\documentclass[12pt,showpacs,showkeys,preprintnumbers,nofootinbib]{revtex4-1}

\usepackage{graphicx}

\usepackage{amssymb,amsmath,graphics,graphicx} \usepackage[usenames,dvipsnames,svgnames,table]{xcolor}
\usepackage{url}

\begin{document}

\title{Random degree-degree correlated networks}

\author{Marlon Ramos$^1$}
\thanks{E-mail: marlon.ramos@fis.puc-rio.br}

\author{Celia Anteneodo $^{1,2}$}
\thanks{E-mail: celia@fis.puc-rio.br}

\affiliation{
$^{1}$Departamento de F\'{\i}sica, PUC-Rio, Rio de Janeiro, Brazil \\
$^{2}$National Institute of Science and Technology for Complex Systems, Rio de Janeiro, Brazil}

\date{\today}

\begin{abstract}
Correlations may affect propagation processes on complex networks. 
To analyze their effect, it is useful to build 
ensembles of networks constrained to have a given value of a structural 
measure, such as the degree-degree correlation $r$, being random in other aspects 
and preserving the degree sequence. This can be done through Monte Carlo optimization procedures. 
Meanwhile, when tuning $r$, other network  properties may concomitantly change. 
Then, in this work  we  analyze, for the $r$-ensembles,  the impact of   $r$ on  properties   
such as transitivity, branching and characteristic path lengths, 
that are relevant when investigating spreading phenomena  on these networks.  
The present analysis is performed for networks with degree distributions of two 
main types: either localized around a typical degree 
(with exponentially bounded asymptotic decay) 
or broadly distributed (with power-law decay).    
Size effects are also investigated. 

\end{abstract}

\pacs{ 
64.60.aq, 
64.60.an, 
05.40.-a  
}

\maketitle

\section{Introduction}
\label{sec:intro}

Complex networks are realistic substrates for simulating 
many social and natural phenomena.
To address the influence of  network topology, 
primarily, different classes of degree distributions $P(k)$ 
can be considered. 
Meanwhile, for a given distribution of degrees,  correlations may   
give rise to important network structure effects on the studied  
process~\cite{vazquez,small,optimizing,epidemic,makse,threshold}. 
These structural effects may have important consequences, for instance, 
correlations may shift the epidemic threshold~\cite{threshold}.  
Although correlation effects may be absent in some cases \cite{absence}, 
in other ones, they can not be neglected.

Despite there are efficient algorithms to generate networks with fixed 
degree-degree correlations \cite{pusch}, real joint probabilities of two or more degrees measured 
in networks of moderate size may be noisy  and hard to be modeled. 
Then, operationally, average nearest-neighbors degree distributions \cite{asymptotic} 
or single quantity measures are used. 
Although other variants have been defined in the literature \cite{mixing,reshuffling}, 
as  quantifier of the tendency of adjacent vertices to have 
similar or dissimilar degrees,     
we will consider the standard measure of (linear) degree-degree correlations,  
namely, the assortativity (Pearson) coefficient  \cite{newman_assor} 
\begin{equation} \label{def_r}
    r=\frac{\langle kk' \rangle_e - \langle k\rangle^2_e  }{\langle k^2 \rangle_e - \langle k\rangle^2_e},  
\end{equation}
where $\langle \cdots \rangle_e$ denotes average over edges and $k$ and $k'$ are 
the degrees of vertices at each end of an edge.   
Despite this coefficient is known to present some drawbacks \cite{cap3}, it is a 
standard and commonly used quantity, hence being worth to be analyzed. 
Moreover, it has the advantage of 
being a single value measure, that is easier to be controlled than other multi-valued quantities.

To analyze the influence of correlations, as well as of any other structural feature, 
 it is useful to build ensembles of networks holding that property,  
while keeping fixed the sequence of degrees. 
As it will be described in Sec. \ref{sec:ensembles}, this kind of ensembles can be 
achieved by means of a suitable  rewiring, 
performed through a standard simulated annealing Monte Carlo (MC) procedure to minimize 
a given energy-like quantity (maximum entropy ensemble approach), 
function of the  graph property to be controlled ($r$ in our case)~\cite{park,foster,noh}.  
Once tuned $r$, it is important to characterize how other 
network properties are altered as by product.
Some interdependencies among certain network properties have already been
numerically shown in the literature, 
for real   as well as for  artificial graphs \cite{foster}. 
Analytical relations have also been derived \cite{estrada,serrano2,dorogovtsev}.
Because of its crucial role in spreading phenomena \cite{newman2001}, 
we will focus here on  the effect of  $r$ over typical distance measures  
as well as on the branching and transitivity of links.

As a measure of the average separation between nodes, 
we consider the average path length \cite{WS}. 
In the subsequent calculations we use the expression,
\begin{equation}
L= \frac{\sum_{i=1}^n \langle L_i \rangle N_i(N_i-1)}{\sum_{i=1}^n N_i(N_i-1)} \, ,
\end{equation} 
where $n$ is the number of (disconnected) clusters and $N_i$ is the number of nodes in cluster $i$.  
Moreover, being $d_{kj}$   the distance (number of edges along the shortest path) 
between nodes $k$ and $j$ (taking $d_{kj}=0$ if the nodes do not belong to the same cluster), 
then 
\begin{equation}
\langle L_i \rangle = \frac{\sum_{j,k=1}^{N_i} d_{kj}}{  N_i(N_i-1)} \,.
\end{equation} 
 
Alternatively, in order to avoid the issue of the divergence of the distance 
between disconnected nodes,  we consider     
the inverse, $1/E$, of the so-called efficiency \cite{latora}  
\begin{equation}
E=\frac{1}{N(N-1)} \sum_{ \substack{1\le i,j \le N \\ i\neq j }}\frac{1}{d_{ij}} \,,
\end{equation}
where $N$ is the number of nodes. 
It represents a harmonic mean instead of the arithmetic one.  
We also compute the diameter $D=\mbox{max}\{d_{ij}\}$.

The transitivity of links can be measured by the clustering coefficient \cite{barrat,newman_clus} 
\begin{equation} \label{ntriangle}
C = \frac{6 n_\triangle}{ \sum_{i=1}^{N} k_i(k_i-1) } \,,
\end{equation} 
where $n_\triangle$ is the number of triangles and $k_i$ is the degree of node $i$.
We also considered the mean value, $\bar{C}$, of the local clustering coefficient $C_i$, 
  defined as $C_i=2 e_i/(k_i(k_i-1))$, where $e_i$ is the number of connections between the 
  neighbors of vertex $i$ \cite{WS}. We took $C_i=0$ when $k_i=0$ or 1. 
Other measures that arise in the decomposition of $r$ \cite{estrada} will also be considered.

Besides detecting interdependencies among structural properties, it is also important to know  
how these properties depend on the system size  $N$.
We will analyze these issues for two main classes of degree distributions 
(Poisson and power-law tailed). 
We will also investigate  real networks degree sequences.

\section{Networks and ensembles}
\label{sec:ensembles}

For each class of networks, we will consider different values of the size, $N$,  
and the mean degree, $\langle k \rangle$, within realistic ranges. 

As a paradigm of the class of networks with a peaked  distribution of degrees,    
with all its moments finite,
we consider the random network of Erd\H{o}s and R\'enyi \cite{ER}. Within this
model, a network with $N$ nodes is assembled by   selecting $M$ different pairs 
of nodes at random and linking each pair. 
The resulting distribution of links is  the Poisson distribution 
$P(k) = e^{-\langle k \rangle}\langle k \rangle^k/k!$, 
where the mean degree is $\langle k \rangle = 2M/N$.

We also analyze networks of the power-law type, i.e., with $P(k)\sim k^{-\gamma}$,  $\gamma >2$,  
corresponding to  a wide distribution of degrees, with power-law tails. 
Then, moments of order  $n\ge \gamma-1$ are divergent.   
We built power-law networks  by means of the configuration model \cite{CM}. 
Following this procedure, one starts by choosing  
$N$ random numbers $k$, drawn from  the degree distribution $P(k)$. 
They represent the number of edges coming out from each node, where these 
edges have one end attached to the node and another still open. 
Second, two open ends are randomly chosen and connected such that, although  
multiple connections are allowed, self connections are not.  
This second stage is repeated until each node attains the connectivity attributed 
in the first step. If eventually  an edge  has an open end, then it is discarded.
However, for large networks, the fraction of discarded edges is negligible.  
To draw the set  of numbers $k$ with probability  $P(k)= {\cal N} k^{-\gamma}$,  
with  $k_{min} \le k \le  k_{max}$ 
(hence the normalization factor is ${\cal N} = 1/\sum_{k_{min}}^{k_{max}} k^{-\gamma}$), 
we used the inverse transform algorithm \cite{transform}. 
Notice that $k_{max}\le N-1$ and $k_{max}>>k_{min}$, 
then we determined $k_{min}$ to fit the selected  value of  $\langle k \rangle$ 
(within a tolerance of at most 1\%), such that
\begin{equation} \label{k1}
\langle k \rangle =\frac{\sum_{k_{min}}^{k_{max}} k^{-\gamma+1}}{\sum_{k_{min}}^{k_{max}} k^{-\gamma}} \simeq 
\frac{\gamma-1}{\gamma-2} \,
\frac{k_{max}^{2-\gamma}- k_{min}^{2-\gamma} }{k_{max}^{1-\gamma}- k_{min}^{1-\gamma}}  
\simeq \frac{\gamma-1}{\gamma-2} k_{min}.
\end{equation}
It is worth mentioning that the value $k=N-1$ is not usually achieved, 
the natural cut-off being $k_c\sim N^\frac{1}{\gamma-1}$ \cite{Dorogovtsev_cutoff}.

In order to attain a desired value of $r$, we follow an standard rewiring 
approach. 
We want to build an ensemble of networks  \{G\} with a given value of   $r$ 
($r$-ensemble) but that are 
maximally random in other aspects, i.e., 
making the fewer number of assumptions as possible about 
the distribution $P(G)$. Then, we use an exponential random graph model, 
such that the set of networks  \{G\} has distribution 
$P(G)\propto {\rm e}^{-H(G)}$, where $H(G)$ is a Hamiltonian or  energy-like quantity  \cite{park}. 
In order to get an $r$-ensemble, with  $r=r_\star$, 
we consider \cite{foster}
\begin{equation} \label{hg}
H(G)=\beta|r-r_\star|\,,
\end{equation}
where $\beta$ is a real parameter. 
The ensemble can be simulated by means of a  MC  procedure: 
at each step, a rewiring attempt is accepted with probability 
$\mbox{min}\{1, e^{-[H(G')-H(G)]}\}$.
Rewiring steps are performed by  randomly selecting two edges that  
connect the vertices  $a$, $b$ and $c$, $d$, respectively, and substituting  those two 
links by new ones connecting $a$, $c$ and $b$, $d$  \cite{rewiring}. 
Movements yielding double links are forbidden. Notice that this process preserves the 
connectivity of each node. 
We start the simulation by taking $\beta=0$ [during at most 100 MC steps (MCS), 
where each MCS corresponds to $N$ attempts]. 
The effect of this stage is basically to destroy multiple edges. 
We did not notice any clear hysteresis effect 
like those observed when controlling, instead, the number of triangles 
with a differente Hamiltonian \cite{hysteresis}. 
Subsequently,  $\beta$ is increased (in increments $\Delta \beta =1000$), at each 50 MCS, 
until $r$ stabilizes, typically attaining the prescribed value $r_\star$.
Then, the quantities of interest are calculated and 
the whole process is repeated, starting with a new degree sequence. 
For power-law degree distributions, we observed that the process is non-ergodic, 
hence we computed sample mean and  standard deviation  over 100 realizations 
of the described protocol. 
We checked that the choice of other expressions for $H(G)$, vanishing at $r_\star$, 
did not significantly affect the results but just the convergence time.

\section{Results}
\label{sec:results}

Let us start by reporting the effects of $r$ on the clustering coefficient $C$.

\begin{figure}[t!]
\includegraphics[width=0.49\textwidth]{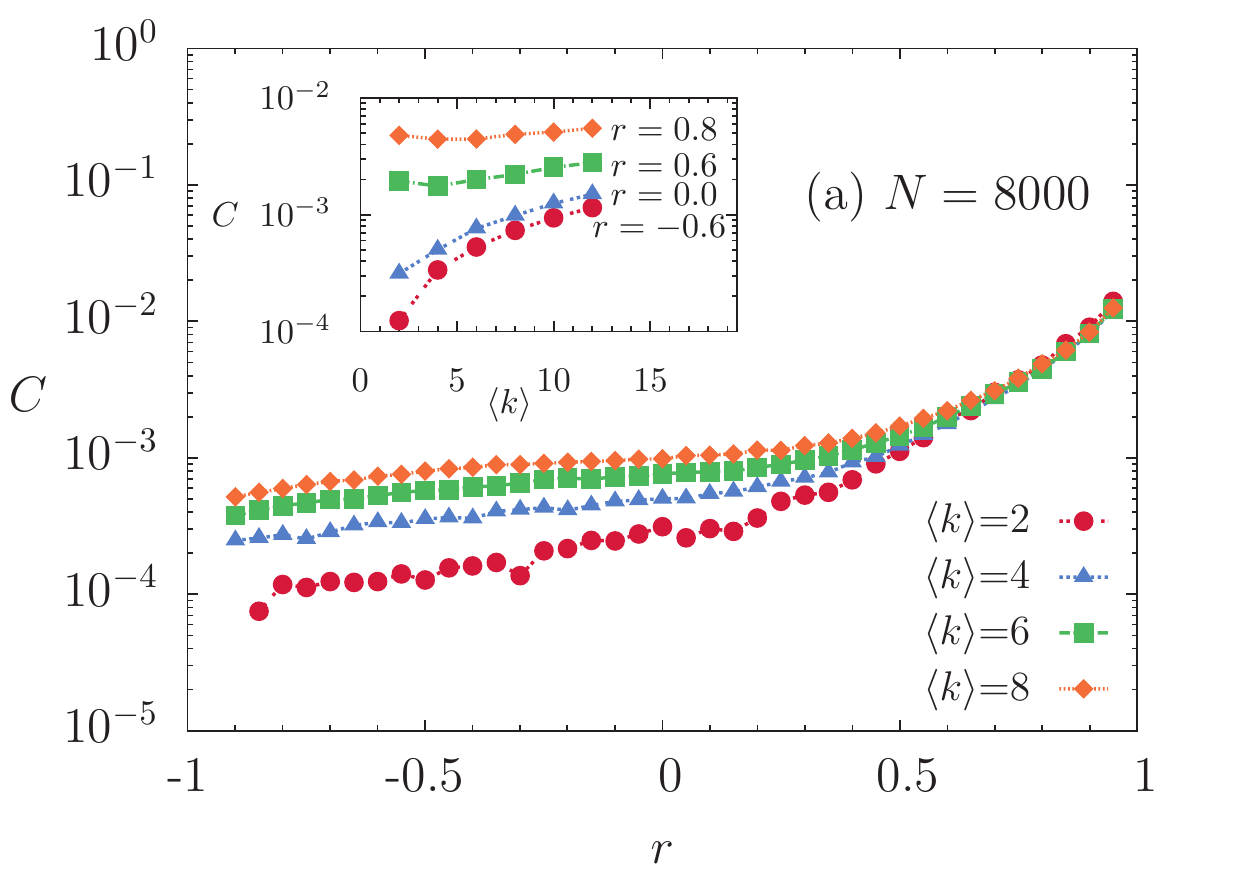}
\includegraphics[width=0.49\textwidth]{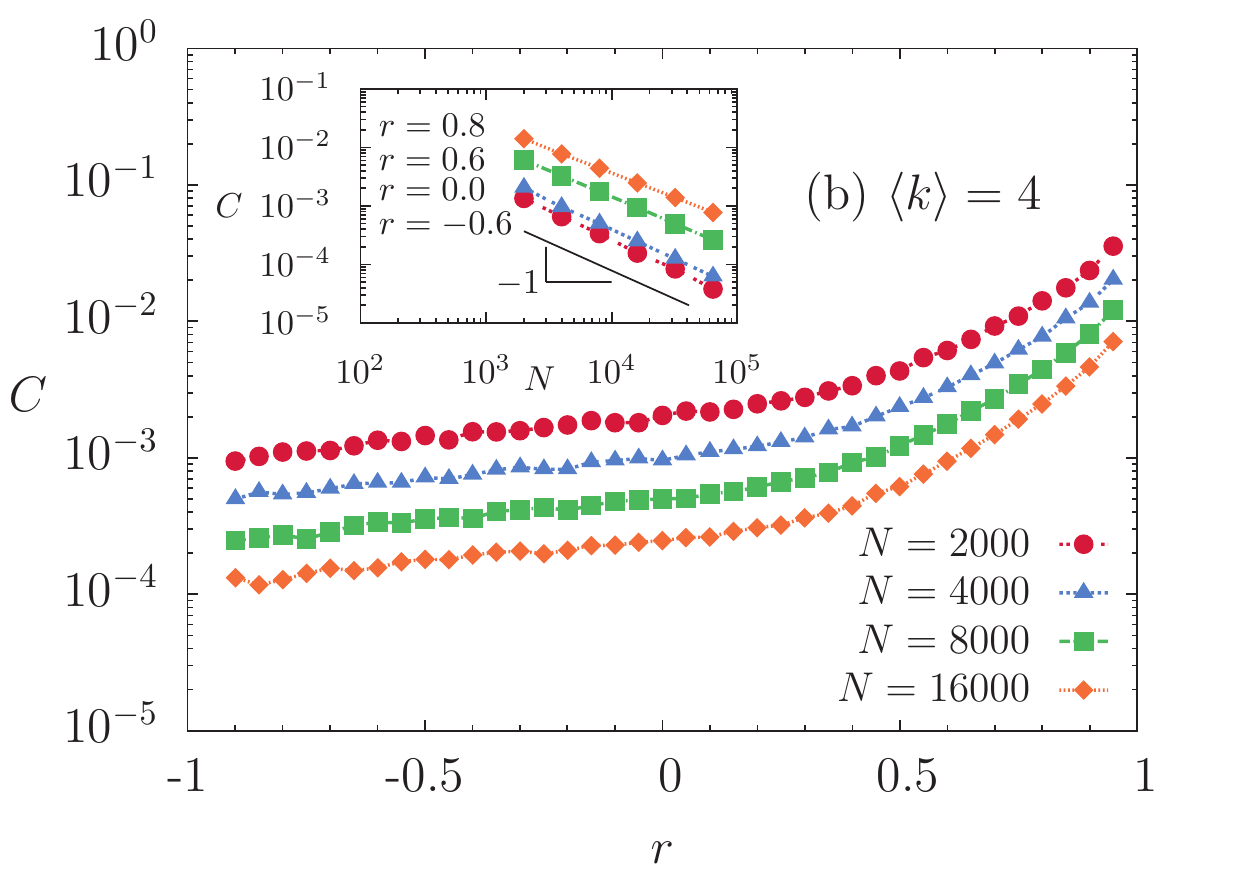} 
\caption{\label{fig:ERC}  
Clustering coefficient $C$ as a function of $r$ for Poisson networks: 
(a) $N=8000$ and different values 
of  $\langle k\rangle$ indicated on the figure.  
The graph shows a monotonic increase of $C$ with $r$. 
There are two regimes: a very slight increase of $C$ with $r$  below $r\simeq 0.5$ and a 
more pronounced one above it.    
(b)  $\langle k\rangle=4$ and different number of nodes $N$, also indicated on 
the figure. As the number of nodes increases, $C$ decays with 
the asymptotic law $C\sim 1/N$, characteristic
of uncorrelated  random graphs. Standard errors are about 10\%.  
Dotted lines are a guide to the eyes. 
The insets show $C$ vs $\langle k \rangle$ (a) and $N$ (b) for selected 
values of $r$ (-0.6, 0.0, 0.6 and 0.8). 
} 
\end{figure}

For the Poisson case, we depict in  Fig.~\ref{fig:ERC}(a) the behavior 
of  $C$ as a function of $r$  for a fixed number of nodes ($N=8000$) 
and different values of the mean connectivity $\langle k\rangle$. 
Very small values of $C$ emerge. The transitivity $C$ monotonically increases with $r$. 
This is consistent with the results of  Ref. \cite{foster} (restricted to $r\ge0$) 
for such kind of networks. 
We observe two regimes with a crossover at $r\simeq 0.5$:  
a very slight increase with $r$  below the crossover  and a 
more pronounced one in the region above it.  
The existence of two regimes could be related to the assymetric character of $r$, which 
does not measure assortativity and disassortativity on the same grounds. 
Below the crossover, $C$ linearly increases with $\langle k \rangle$ about one order of 
magnitude within the 
analyzed range. 
Meanwhile, above the crossover, 
$C$ remains of the same order when the average connectivity increases, even for 
small  $\langle k \rangle$  (also see the inset of  Fig.~\ref{fig:ERC}(a) 
where $C$ is plotted vs $\langle k \rangle$ for selected 
values of $r$). In Sec. \ref{sec:final}, we will discuss these issues in more detail.
For the mean local clustering coefficient $\bar{C}$, 
we obtained a qualitatively   similar dependence on $r$ than that observed for the clustering coefficient $C$. However, the increase of $\bar{C}$ with  $\langle k \rangle$ is linear for any fixed $r$. For $r=0$, $C=\bar{C}=\langle k\rangle/N$, as expected.

In  Fig.~\ref{fig:ERC}(b), size effects are exhibited for $\langle k\rangle=4$, 
representative of the other values considered. 
As the number of nodes increases, $C$ decays as
$C\sim 1/N$ for all $r$ (as depicted in the inset). 
Therefore, in the $r$-ensemble of Poisson networks, 
transitivity is only a finite-size effect and 
vanishes in the infinite network (thermodynamic) limit with 
the same asymptotic law $C\sim 1/N$ that
for an  uncorrelated  random graph \cite{APC}.

\begin{figure}[b!]
\includegraphics[width=0.49\textwidth]{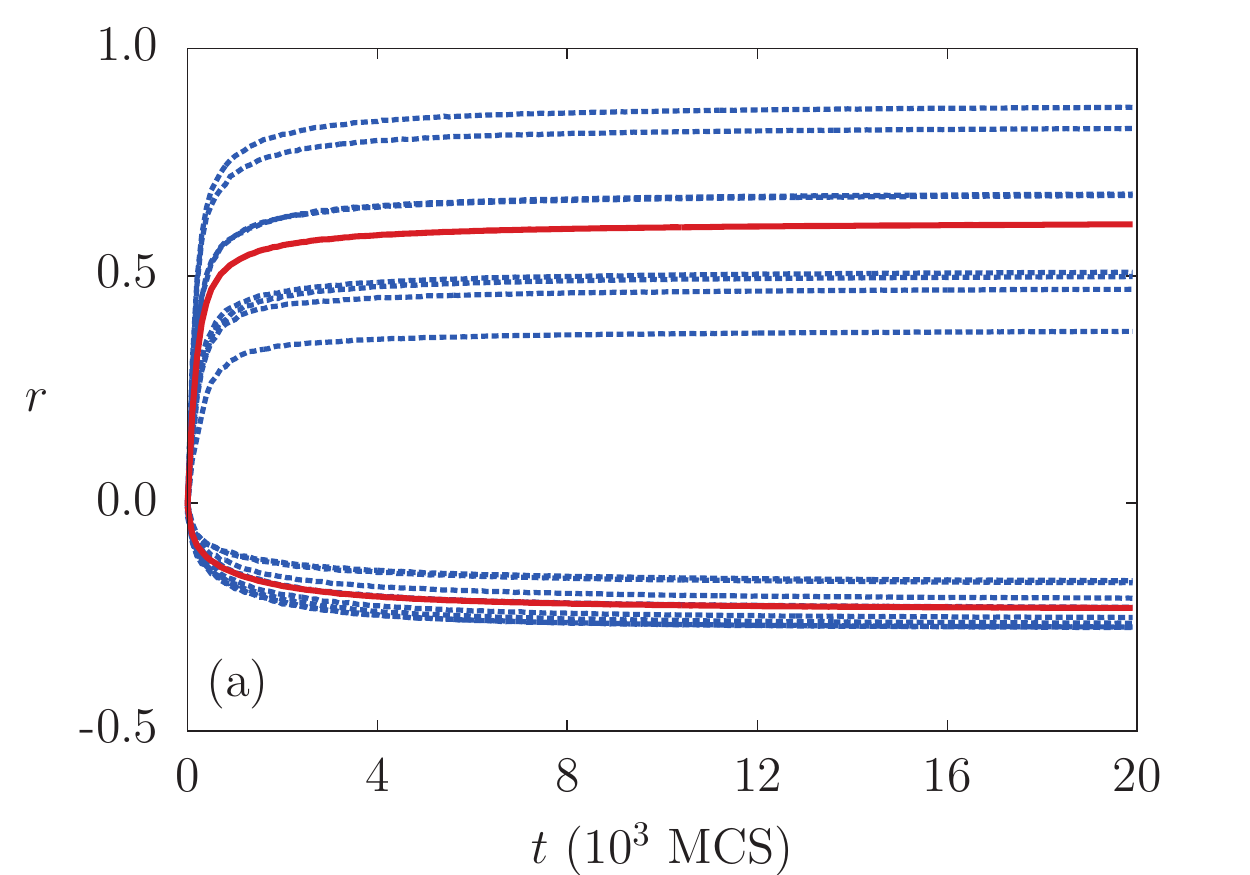}
\includegraphics[width=0.49\textwidth]{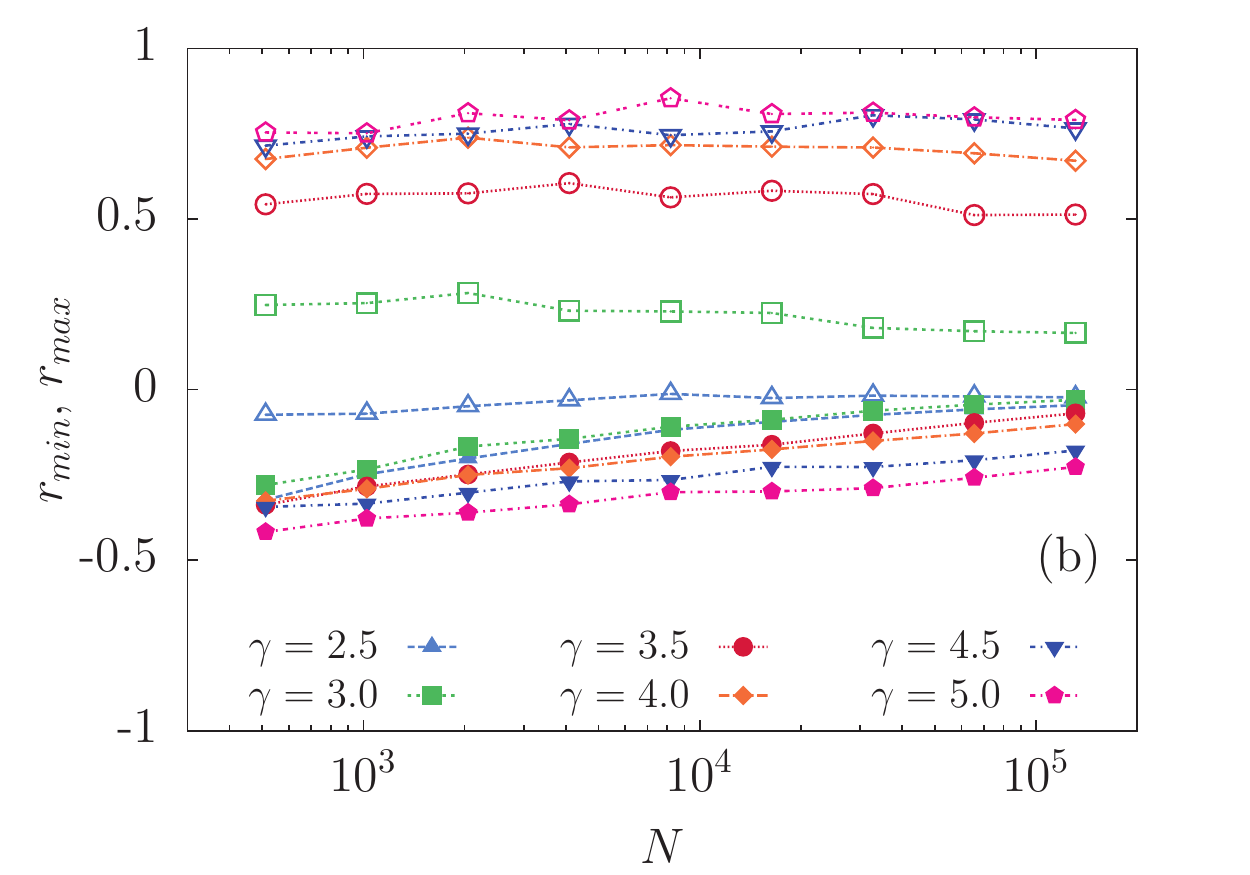} 
\caption{\label{fig:rlim}  
The range of allowed values of $r$ is restricted for the power-law class:   
(a) Time evolution of $r$, after setting $r_\star=1$ (-1)  to obtain $r_{max}$ ($r_{min}$), 
for networks with power-law degree distribution (with $\gamma=3.5$ and $N=5000$). 
Shown are 8 individual samples (thin lines) and their respective averages (thick lines).  
(b) Average extreme values [$r_{max}$ (open symbols) and $r_{min}$ (filled symbols),  
standard errors are at most 50\%] vs   
system size $N$ for different values of $\gamma$ indicated on the figure. 
Dotted and dashed lines are guides to the eye for $r_{max}$ and $r_{min}$, respectively. 
In all cases $\langle k \rangle = 4.00 \pm 0.04$. 
For a given size, the allowed interval of $r$ is narrower for lower $\gamma$. 
} 
\end{figure}

For the power-law class, the range of allowed values of $r$ is restricted. 
That is, values of $r$ arbitrarily different from zero can not be attained in 
typical realizations of the MC protocol described in Sec. \ref{sec:ensembles}. 
In order to determine the typical maximal (minimal) values, 
$r_{max}$ ($r_{min}$), 
we imposed $r_\star= 1$ (-1) and detected the stationary values of $r$. 
The time evolution of $r$ for $r_\star= 1$ (-1) is illustrated in Fig.~\ref{fig:rlim}(a) 
for $\gamma=3.5$, $N=5000$ and $\langle k \rangle \simeq 4$. 
Notice the large deviations amongst the steady values of different realizations mainly for the upper bound. 
We verified that this picture does not change by implementing other definitions of $H(G)$ in Eq.~(\ref{hg}), 
e.g., $\beta|r-r_\star|^\alpha$, with $\alpha \neq 1$. 
Average extreme values (over 100 samples, after $3\times 10^4$ MCs) 
are displayed  in Fig.~\ref{fig:rlim}(b),  
as a function of the system size, for different values of $\gamma$.  
For fixed size, the lower $\gamma$, the narrower the allowed interval of $r$.
In fact, in networks constrained to a given degree sequence,  
structural limitations (or correlations) arise: 
either   multiple connections or dissortative two vertices correlations \cite{newman03}.
For instance, the exclusion of multiple connections hampers the natural tendency that hubs connect among them, 
hence diminishing the assortativity. This effect is more pronounced the smallest $\gamma$.  
For fixed $\gamma$, the interval   shrinks with system size, for $\gamma<3$, due to the divergence of fluctuations in the large $N$ limit \cite{cap3}.
In Ref. \cite{asymptotic}, similar restriction was also observed  
for $2<\gamma <3$, although instead of the average connectivity, 
$k_{min}$ was kept constant ($k_{min}=6$). 
In that case, it was reported that the upper and lower bounds 
both tend to zero, hence $r\to 0$ in the infinite network limit. 
In fact, we observe that in that interval of $\gamma$ (e.g., $\gamma=2.5$) 
both bounds are negative and as $N$ increases the allowed 
interval collapses to a negative value that tends to zero. 
We noticed restriction in the correlation bounds for $\gamma>3$ too. 
As $N$ grows, the lower bound also increases towards zero 
or at least to a small finite value. 
Simultaneously, the upper bound seems more stable, 
however its asymptotic behavior is not neat yet, even having considered up to $N>10^5$. Moreover, as $N$ increases, it takes longer to attain steady states.

The allowed interval of $r$ is quite restricted for scale-free networks. 
However, we still analyzed systematically cases with $\gamma>3$ 
($\gamma=3.5$, $4.0$ and $4.5$), yielding finite second moment. 
Even, in this range, the accessible interval of $r$ is limited, 
then, we proceeded as follows. 
If the desired $r_\star$ is not attained, 
within a tolerance of $10^{-3}$, in $2\times 10^4$ MCS, that instance is 
discarded and we make a new trial. 
If we did not attain 100 successes  in 200 trials, the procedure is interrupted.   
Alternatively to the configuration model, we also started from networks generated by 
preferential attachment \cite{PA}, yielding similar results.

\begin{figure}[h!]
\includegraphics[width=0.49\textwidth]{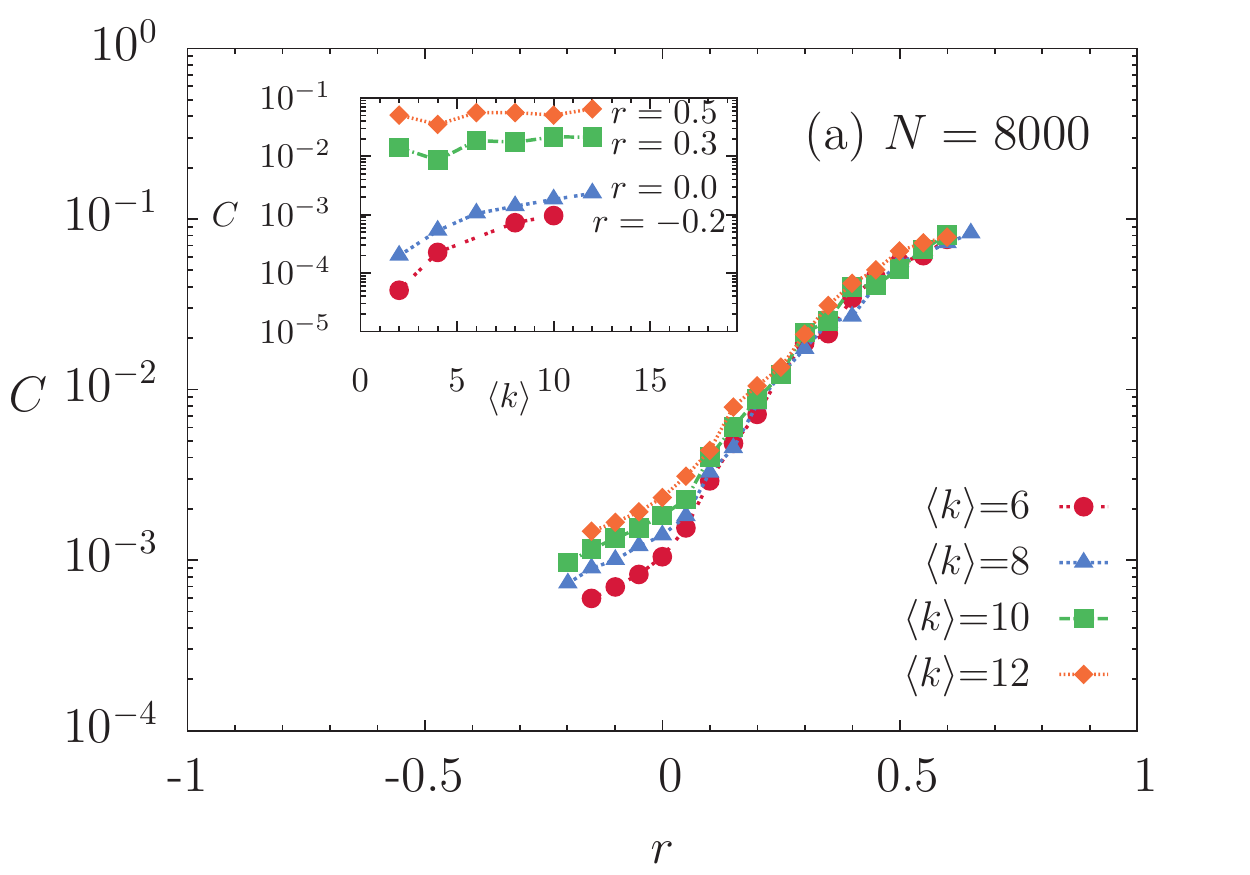}
\includegraphics[width=0.49\textwidth]{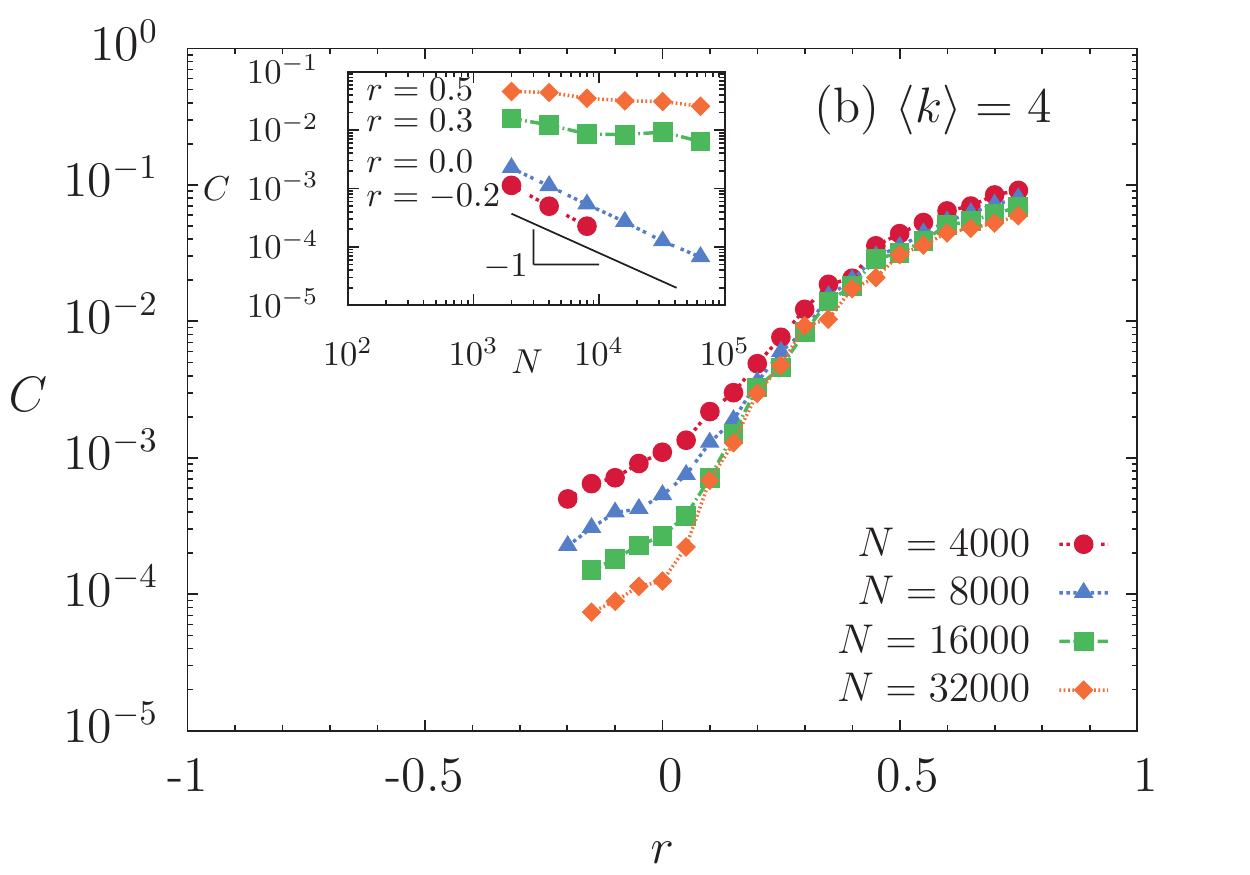} 
\caption{\label{fig:SFC}  
Clustering coefficient $C$ as a function of $r$, as in 
Fig.~\ref{fig:ERC} but for power-law networks with $\gamma=4.0$: 
(a) $N=8000$ and different values  of  $\langle k\rangle$. 
In the assortative region, $C$ reaches larger values than in the Poisson class. 
(b) $\langle k\rangle=4$ and different number of nodes $N$. 
Note that for  assortative networks a finite degree of clustering seems to persist for large networks.
Standard errors reach 50\% for the lowest values of $C$.
In the inset, missing values are due to the limitation in attaining the 
prescribed values of $r$.  
} 
\end{figure}

The outcomes for the power-law class with $\gamma=4.0$ are displayed in Fig.~\ref{fig:SFC}. 
Standard errors are larger  in the power-law case, likely due to the variability in the tails 
of the distribution of links from sample to sample. 
Outcomes for the other two values of $\gamma$ studied  (3.5 and 4.5) 
display features similar to those of the case $\gamma=4.0$ used 
as illustrative example, despite the third moment becoming divergent at $\gamma=4.0$.  
Two regimes are also observed,   with the crossover now closer to $r=0$,  
but some qualitative differences appear in comparison to the Poisson case. 
$C$ rapidly increases with $r$, 
attaining,  for assortative networks,    larger values than in the Poisson class.  
These large values are due to the inclusion of highly connected nodes, absent in the Poisson  
networks, that for large $r_\star$ tend to gain links among them, contributing 
strongly to $r$ and also to $C$.

With respect to finite size effects, below the crossover 
the small non-null $C$ is again due only to the finite size of the network. 
However, in the assortative region (above the crossover), 
it seems that a finite degree of clustering persists for large networks 
(see inset of Fig.~\ref{fig:SFC}(b)), in contrast to the Poisson case and to the 
dissassortative region.  
In fact, notice that, when $N$
increases one order of magnitude, $C$ decreases also one order of magnitude  
in the dissortative region, while $C$ remains of the same order  in the  assortative interval. 
Even if $C$ vanished  in the infinite size limit, since the decay is very slow, 
then  an  effective  clustering would remain in moderate, or even large, size networks. 
We will discuss the interplay between $C$ and $r$ further in Sec. \ref{sec:final}.
For the mean local clustering coefficient $\bar{C}$, 
a qualitatively   similar dependence on $r$  is observed, 
but with smaller values. 
Moreover, $\bar{C}$ increases linearly with  $\langle k \rangle$, 
in the analyzed range, for any fixed $r$, not only for dissortative networks, 
and  $\bar{C}$ decays with $N$ for any $r$. For$r=0$, $C=\bar{C}=(\langle k^2\rangle - \langle k\rangle)^2/(N\langle k \rangle^3)$ \cite{dorogovtsev}, as expected.

\begin{figure}[h!]
\includegraphics[width=0.52\textwidth]{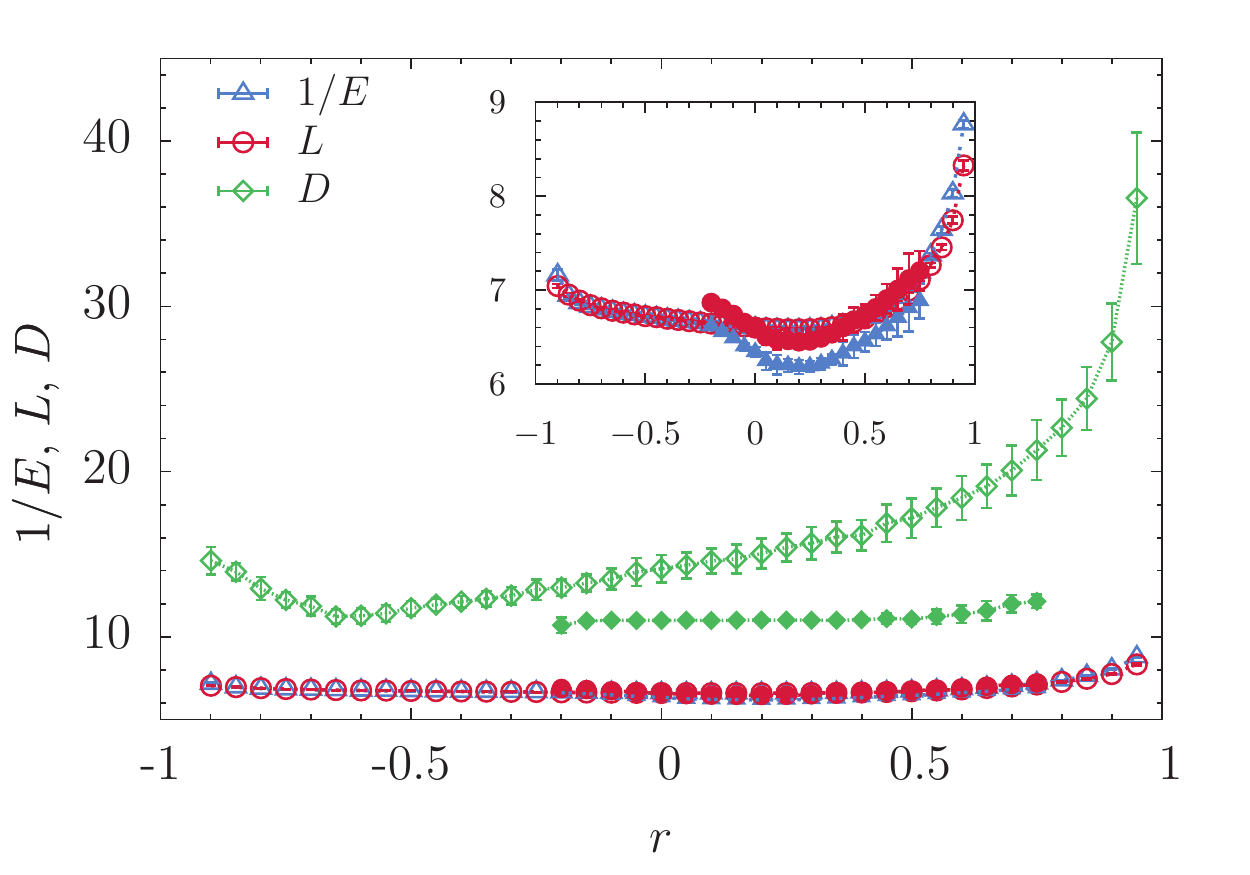}\caption{\label{fig:distances}  
Distance measures $1/E$, $L$ and diameter $D$, as a function of $r$ for   
Poisson (open symbols) and $\gamma=4.0$ power-law  (filled symbols) networks. 
In all cases $N=8000$ and   $\langle k\rangle =4.00 \pm 0.04$. 
The inset is a zoom of the main plot. At first sight,  
the two types of network display similar values of  $1/E$ and $L$,  
for a given $N$ and $\langle k \rangle$. 
The diameter is more sensitive to the type of network and is more 
affected by $r$ in the Poisson case.
} 
\end{figure}

Let us analyze now the influence of $r$ on  network characteristic lenghts. 
The dependency of the measures  $1/E$, $L$ and $D$ on $r$ is depicted in 
Fig.~\ref{fig:distances}, for Poisson and power-law distributed networks, with  $N=8000$ and $<k>\simeq 4$.
$1/E$ and $L$ have close values, systematically shifted. 
In first approximation both types of network yield similar values of  $1/E$ (hence also $L$),  
given $N$ and $\langle k \rangle$.  
However, the diameter $D$ is more dependent on the type of network. It is larger and 
is more strongly affected by $r$ in the homogeneous Poisson case.

\begin{figure}[t!]
\includegraphics[width=0.49\textwidth]{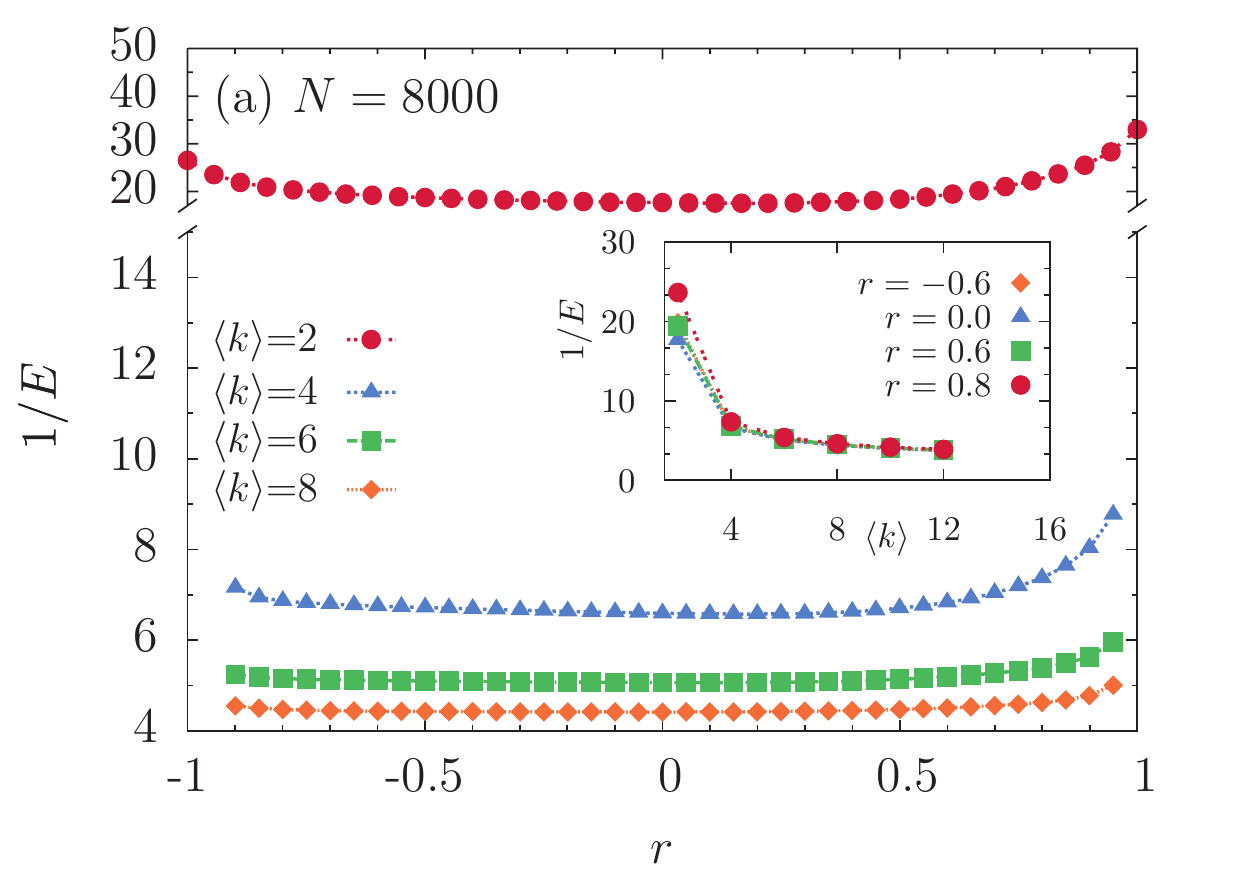}
\includegraphics[width=0.49\textwidth]{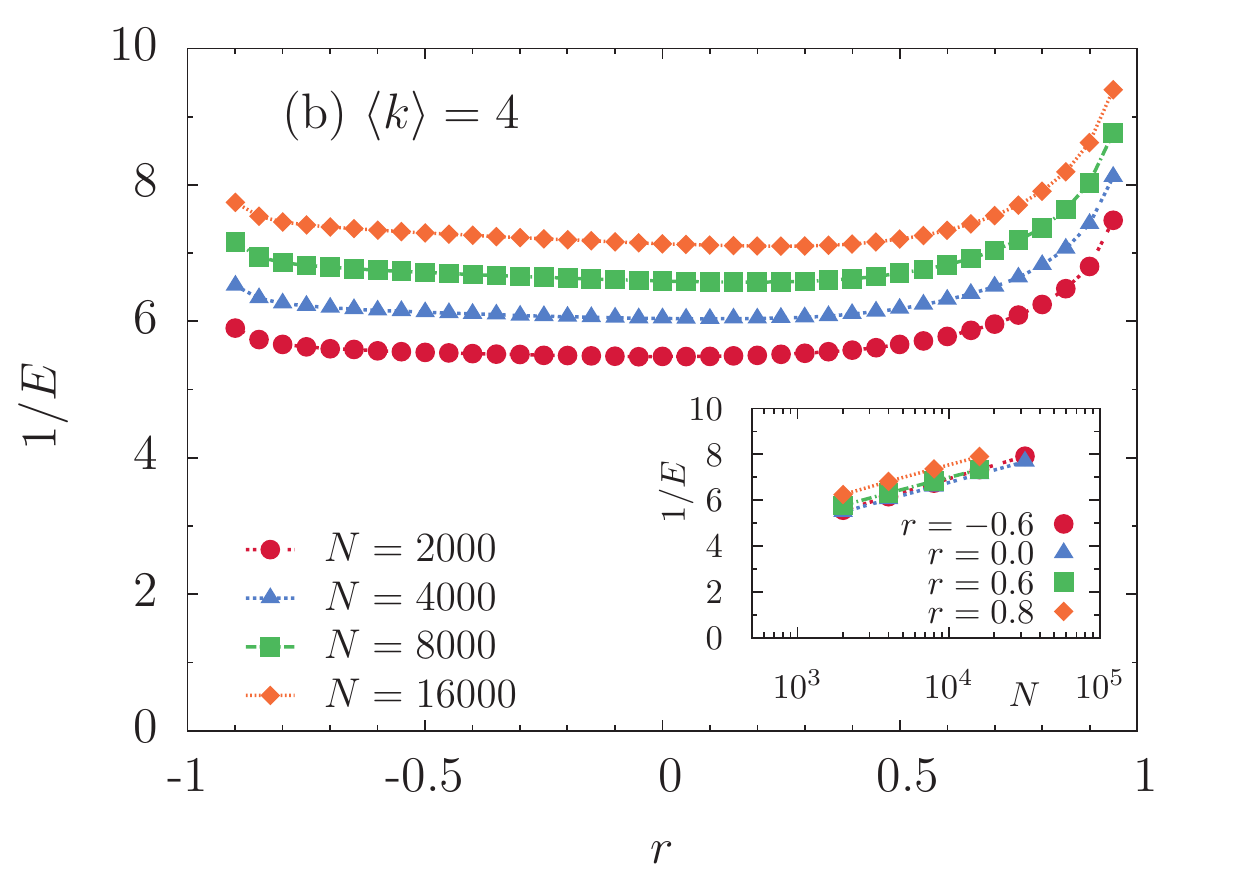} 
\caption{\label{fig:ERL}  
Mean distance  $1/E$ as a function of $r$ for Poisson networks: 
(a) $N=8000$ and different values 
of  $\langle k\rangle$ indicated on the figure. 
The networks exhibit the small-world property.
(b)  $\langle k\rangle=4.00 \pm 0.04$ and different number of nodes $N$ indicated on 
the figure.  The effects of $r$ on the mean path are significant only 
for small $\langle k \rangle$ because of the fragmentation of the network.} 
\end{figure}

\begin{figure}[t!]
\includegraphics[width=0.49\textwidth]{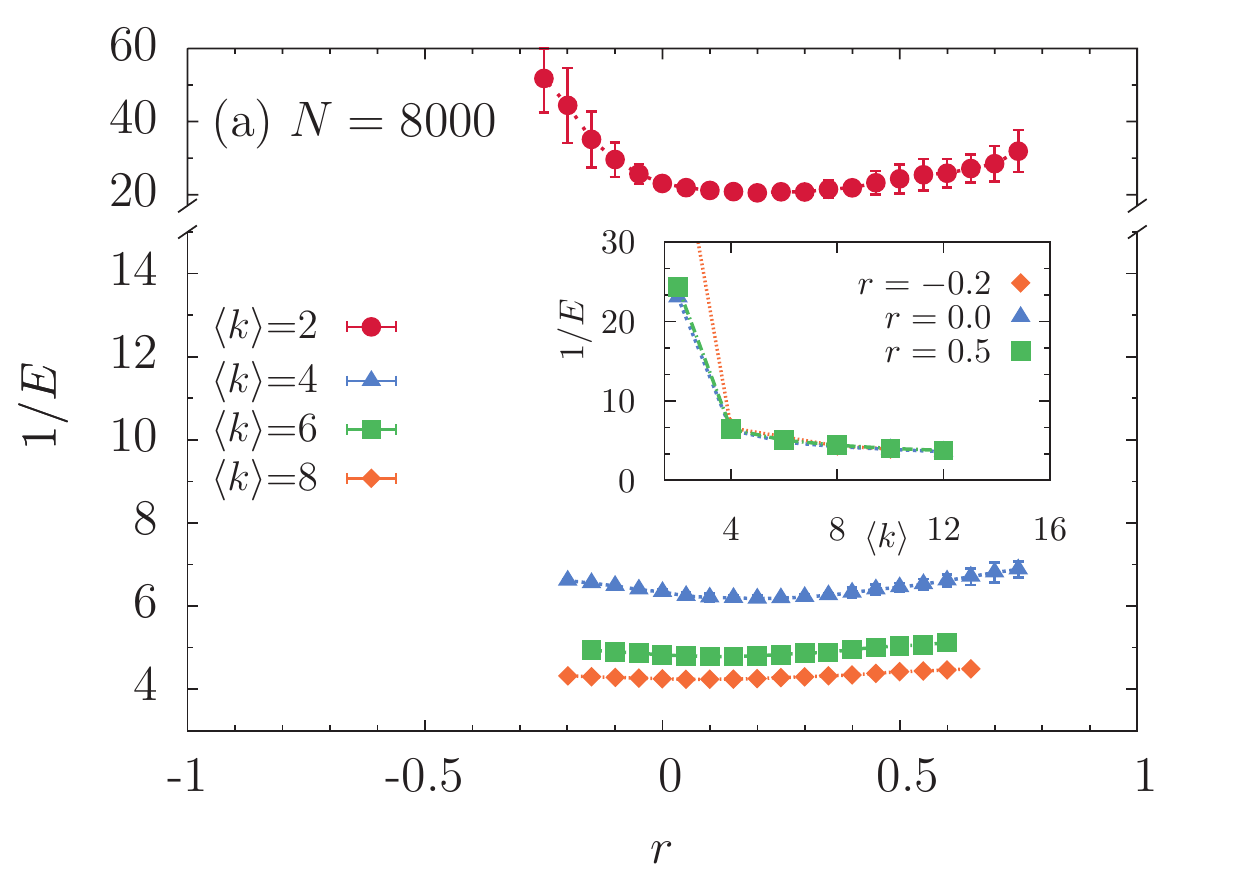}
\includegraphics[width=0.49\textwidth]{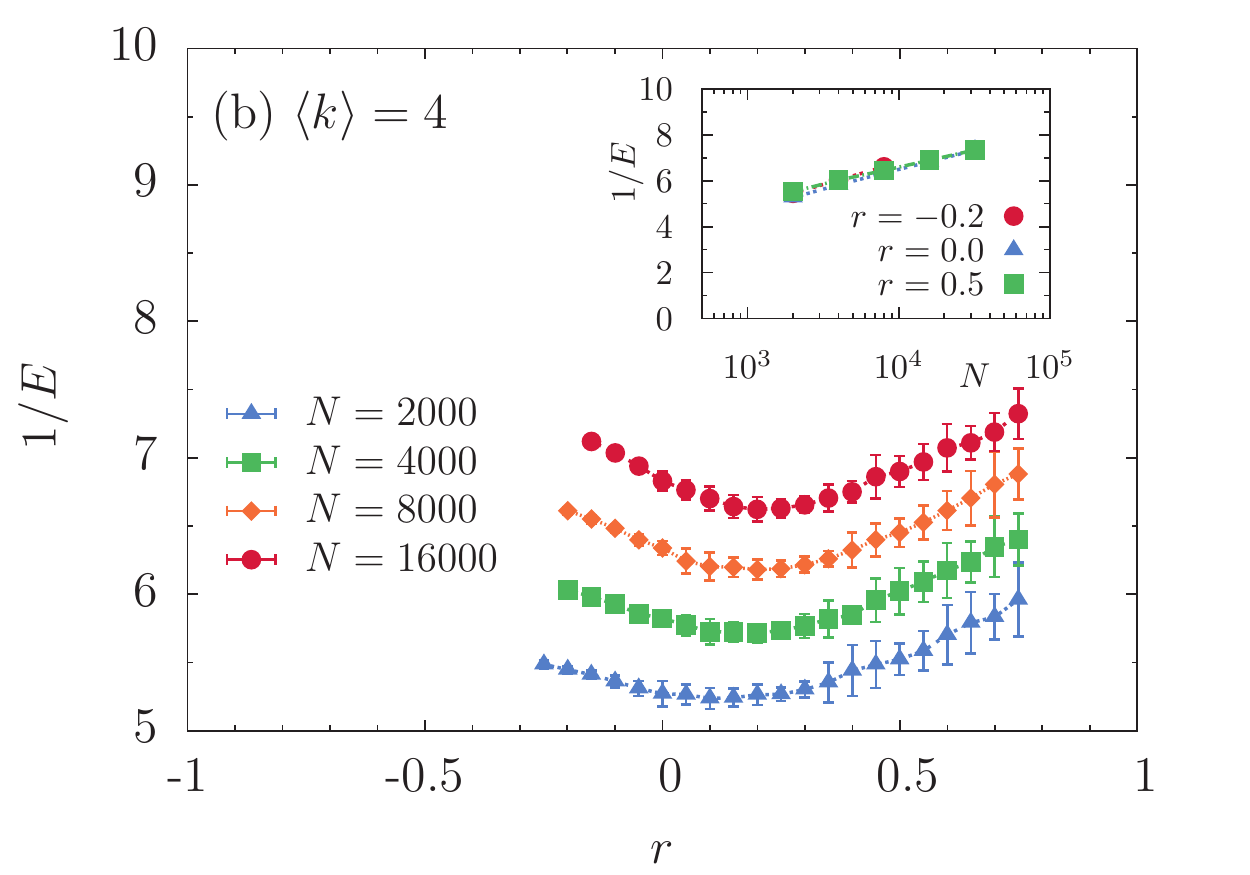} 
\caption{\label{fig:SFL}  
Mean distance  $1/E$ as a function of $r$ as in 
Fig.~\ref{fig:ERL} but for power-law network with $\gamma=4.0$. 
In this case, paths are shorter  than in the Poisson nets. }
\end{figure}

Plots of $1/E$ vs $r$ for different values of $N$ and $\langle k\rangle$ 
are shown in Figs. \ref{fig:ERL} and \ref{fig:SFL} for Poisson and power-law networks 
respectively. 
In both cases, the networks display the small-world property \cite{WS}  
(even smaller in the power-law case) with a slow (logarithmic)  
increase with $N$ and a smooth decrease with $\langle k \rangle$ (see insets of 
Figs. \ref{fig:ERL} and \ref{fig:SFL}).

In the Poisson case,  
the mean path does not depend on $r$ significantly when $\langle k \rangle$ is not too small ($\ge 6$), 
as indicated by the relatively flat plots in Fig.~\ref{fig:ERL}(a). 
Only for small $\langle k \rangle$ there are important effects for assortative correlations. 
For instance, for $\langle k \rangle \simeq 4$ (Fig.~\ref{fig:ERL}(b)),  
$1/E$ increases in about two units from $r\simeq 0$ to $r\simeq 1$, for all the analyzed sizes.  
This effect is still larger for $\langle k \rangle=2$ where $L$ increases by a factor about two  
in the same interval of $r$, as shown in Fig.~\ref{fig:ERL}(a) for $N=8000$.   
 
\begin{figure}[h!]
\includegraphics[width=0.52\textwidth]{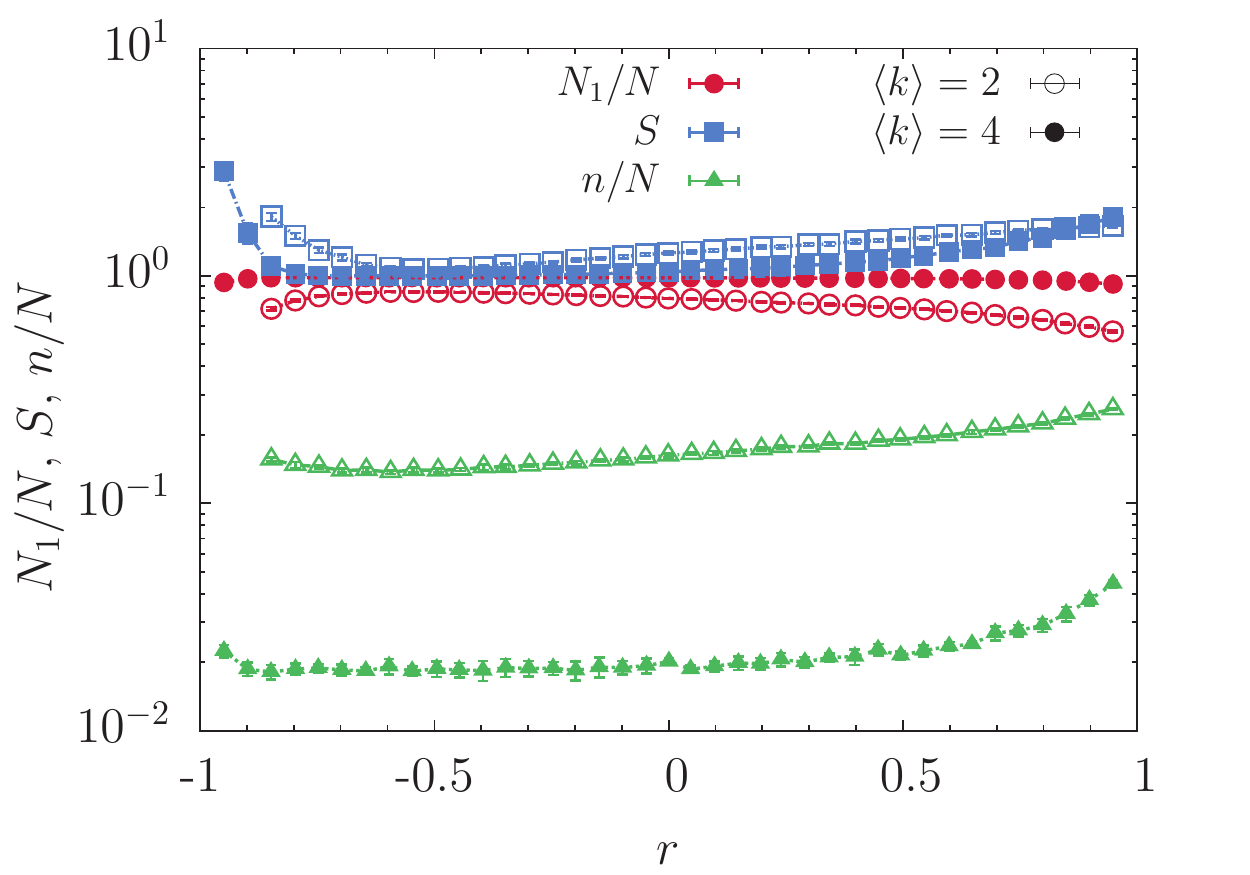}
\caption{\label{fig:clusters}  
Clusters analysis. Plots of $N_1/N$ (where $N_1$ is the size of the largest cluster) [circles], 
average size of finite size clusters, $S$ [squares], and number of clusters, $n$ [triangles], 
as a function of $r$ 
for Poisson networks with  $\langle k\rangle=2$ (open symbols) and 4 (filled symbols). 
The fragmentation of the network observed for high values of the assortativity reflects 
the grouping of vertices with the same degree.  
The figure shows outcomes for $N=8000$. Outcomes for sizes  $N=4000$, 8000 and 16000   
all collapse into single curves (not shown). 
} 
\end{figure}

In order to further interpret these results, we investigated the cluster structure of the resulting 
rewired networks. We measured the size of the largest cluster (let us call it $N_1$), 
the number $n$ of clusters,  and the average size $S$ of the clusters different from the largest one. 
The plots are presented in Fig.~\ref{fig:clusters} for  $\langle k \rangle=2$ and 4. 
For   $\langle k \rangle=4$, the relative size of the largest cluster (circles) is about 0.98 for most of the 
range of $r$, notice however that it slightly decays towards $|r|=1$  
(which is  more evident for $\langle k \rangle=2$). 
As $\langle k \rangle$ increases, the number of fragments 
rapidly decays and the average size $S$ (triangles) tends to unit, 
meaning that only single nodes are disconnected 
(recall also that  $P(0)={\rm e}^{-\langle k \rangle}$).  
Therefore the increase of the mean distance towards $|r|=1$, observed for low 
$\langle k \rangle$  in Fig.~\ref{fig:ERL}, 
may  simply reflect the fragmentation of the network.
Clearly, for high values of the assortativity, 
the network tends to fragment into groups of vertices that have the same degree.   
Moreover, for  values of $\langle k\rangle$  approximately larger than 1, 
the percolation analysis performed by Noh on Poisson networks \cite{noh}  shows that 
the size of the largest cluster is smaller for assortative networks than for  dissortative and neutral ones.
Meanwhile, as  $\langle k \rangle$ increases, 
 the fraction of vertices that do not belong 
to the largest cluster becomes negligible, although more slowly the more assortative the network.
Therefore, in such large $\langle k \rangle$  limit, 
the mean distance remains invariant under changes of $r$. 
Hence, transport processes modeled in these networks may suffer important impact 
when $r$ is large and $\langle k \rangle$ small. 
The longer the typical separation between nodes, the slower the propagation.

In  power-law networks, for fixed $N$ and $\langle k\rangle>2$, there is an interval of $r$ 
where paths are shorter  than in the Poisson nets (Fig.~\ref{fig:distances}), 
and still shorter as $\gamma$ decreases (not shown). 
Moreover $1/E$ becomes more sensitive to the coefficient $r$ (smile shape), 
in the region where plots are flat for Poisson networks. 
Notice also that minimal mean paths occur for slightly assortative correlations ($r \gtrsim 0$), 
slowly increasing with $N$ (Fig.~\ref{fig:SFL}(b)). 
The analysis of clusters for $\gamma=4.0$, shows that for $\langle k\rangle \ge 4$ 
there is a single cluster of size $N$, for all $r$. 
Only for $\langle k\rangle=2.0$ we observed fragmentation with  
$N_1/N\simeq 0.7-0.8$, $n/N\simeq 0.06$, $S\simeq 4$ for all $N>2000$ (plots not shown).

For the mean path $L$, we observed qualitatively 
similar outcomes although shifted to slightly higher values, as illustrated in 
Fig.~\ref{fig:distances}.

\begin{figure}[t!]
\includegraphics[width=0.49\textwidth]{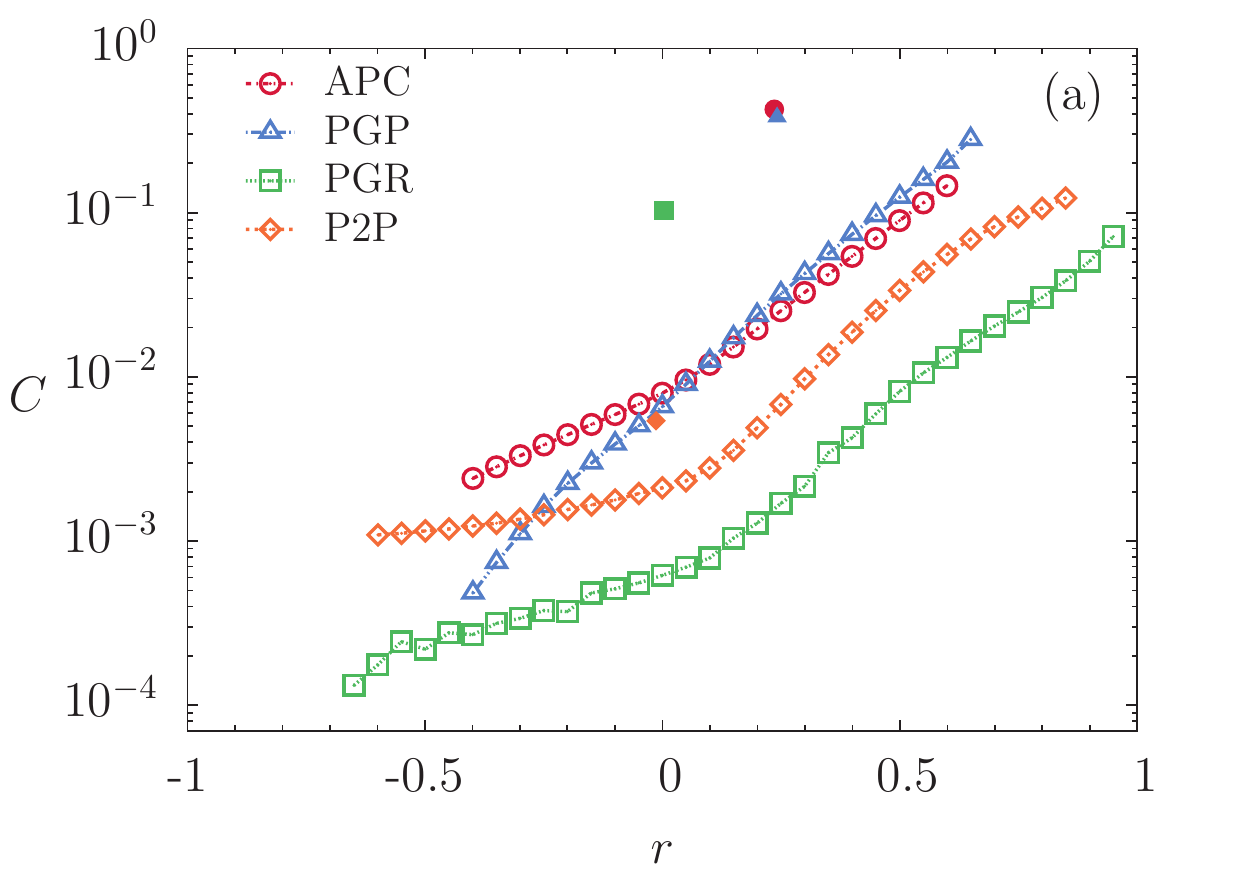}
\includegraphics[width=0.49\textwidth]{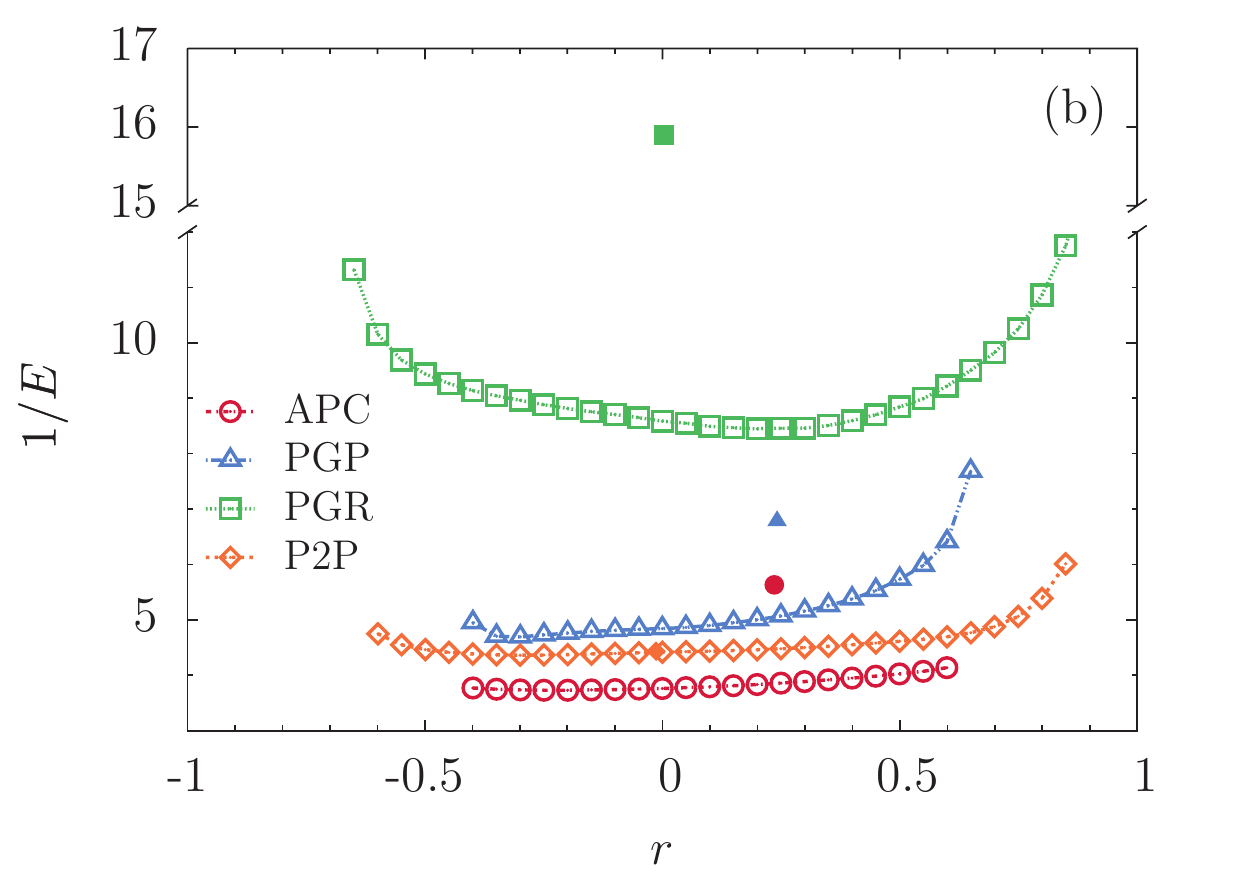}
\caption{\label{fig:real}  
$C$ and $1/E$ vs  $r$, for real networks. Original values, before rewiring, 
are also indicated (filled symbols). PGP (Pretty Good Privacy encrypted communication network) 
\cite{PGP}:  $N=10680$, $\langle k\rangle\simeq 4.55$; P2P (Gnutella peer-to-peer network) 
\cite{P2P}: $N=10876$, $\langle k\rangle\simeq 7.35$; PGR (power grid) \cite{PGR},  
$N=4941$, $\langle k\rangle\simeq 2.67$; APC (astrophysics collaboration) 
\cite{APC}: $N=16706$, $\langle k\rangle\simeq 14.5$. 
Besides the different details of real degree sequences, 
we can interpret the main features of these nets in terms 
of those observed for the Poisson and power-law classes. 
} 
\end{figure}

We also applied the rewiring procedure described in Sec. \ref{sec:ensembles} to 
real degree sequences. Networks were symmetrized and edge weights   were ignored. 
In Fig.~\ref{fig:real} we depict the behavior of $C$ and $1/E$ vs  
$r$, for 
the PGP (encrypted communication network using Pretty Good Privacy encryption algorithm) 
largest component \cite{PGP}, 
P2P (Gnutella peer-to-peer network) \cite{P2P}, 
PGR (electrical power grid of the western United States), example of small-world network \cite{PGR} and 
APC (astrophysics collaboration network) \cite{APC}. 
First notice that in all cases the clustering $C$ is much larger in real networks than in 
the randomized ($r$-ensemble) ones, as already observed for other examples in Ref. \cite{foster}. 
The mean distance is also typically larger in the real networks. 
An exception is  P2P network, characterized by a value of $1/E$ typical of the $r$-ensemble.
Rewired real networks display some of the typical behaviors observed for the artificial cases. 
Let us make some remarks arising from comparisons. 
(i) PGR (power grid) displays plots of $C$ vs $r$ and $1/E$ vs $r$ that are in good accord with 
those observed for  similar parameters $\langle k \rangle$ and $N$ of the Poisson case. 
In fact its degree distribution decays exponentially.    
(ii) P2P presents power-law decay of the degree distributions, for $k>10$, with  exponent close to 
$\gamma =4$. Both plots of P2P are in agreement with those obtained 
for the $\gamma=4$ class with similar values of $N$ and $\langle k \rangle$, despite the 
distributions only share in common the power-law tail. 
(iii) PGP (of size similar to P2P) has a degree distribution that decays 
with exponent $\gamma<3$ for $k<50$ and $\gamma\simeq 4$ beyond \cite{PGP}. 
The plot for   $1/E$ vs $r$  presents  larger values of $1/E$ than P2P consistent with 
its $\langle k \rangle$. However, the plot $C$ vs $r$ of PGP deviates 
from the standard behavior, presenting larger values of $C$ that increase with $r$ in a single regime. 
The interval of allowed values of $r$ is sensitive also to features other than the tails.  
These deviations  can be attributed to different initial power law regimes.   
(iv) Finally, APC has a power-law decay with exponent $\gamma \simeq 1$ and exponential cut-off for 
$k>50$ \cite{APC}. The low and constant plot of $1/E$ vs $r$ is expected for a network with large 
$\langle k \rangle$, almost independently of the class of degree distribution. The large values 
of $C$ are also consistent with heterogeneous distributions with large $\langle k \rangle$. 

Then, despite the different details of real degree distributions, the main  observed features 
can be interpreted in terms of the analyzed classes with corresponding values of parameters 
$\langle k \rangle$ and $N$.

\section{Discussion and final remarks}
\label{sec:final}

For all classes of networks considered, $C$ increases with $r$  in the whole allowed range of $r$. 
This behavior has already been observed in Ref.  \cite{foster}, 
where only positive values of $r$ were analyzed 
and also in Ref. \cite{serrano2} although different definitions of clustering and correlation 
were used. 
However, we observed that, in the $r$-ensemble,  a non-vanishing clustering coefficient $C$  
is typically due to finite size effects, such that,   
in the large size limit,   network transitivity vanishes as $1/N$. 
In contrast, for  power-law networks characterized by $r$ above a threshold,  
a significantly non-null transitivity arises, apparently persisting for large $N$.  

In any case, since rewiring in the $r$-ensemble turns $C$ typically small,  
transitivity does not  
seem to contribute for attaining the prescribed value of $r$.
To identify the  factors that contribute to $r$, it is useful to rewrite 
Eq.~(\ref{def_r}). 
Recalling that $\langle k^n\rangle_e=\langle k^{n+1}\rangle/\langle k\rangle$ \cite{mendes}, 
where $\langle \cdots \rangle$ (without subindex) means  computed over the degree distribution $P(k)$, 
then  Eq.~(\ref{def_r}) becomes
\begin{equation} \label{r1}
r=\frac{ \langle k \rangle^2  \langle k k' \rangle_e  -\langle k^2\rangle^2}
{\langle k\rangle \langle k^3\rangle -\langle k^2 \rangle^2} \,.
\end{equation}

Following the decomposition made by Estrada \cite{estrada}, notice that  the quantity 
$\tilde{P}_3\equiv\sum_{(k,k')}(k-1) (k'-1)$, 
where the sum is performed over all the different pairs of neighboring vertices, 
is the number of paths of length three, then $\tilde{P}_3=3 n_\triangle + P_3$, 
where  $P_3$ is the number of nontriangular paths of length three (involving four vertices). 
As done in Eq.~(\ref{ntriangle}) for $3 n_\triangle$, let us scale  also $P_3$   by the number of wedges 
(paths of length two) $P_2=\frac{1}{2}\sum_i k_i(k_i-1)$, defining $B=P_3/P_2$ (scaled branching) \cite{estrada}. 
Then Eq.~(\ref{r1}) can be written as
\begin{equation} \label{r2}
r=\frac{ \langle k \rangle (\langle k^2 \rangle-\langle k \rangle) 
\left(  B+1- \frac{\langle k^2\rangle}{\langle k\rangle}  +C \right) }
{\langle k\rangle \langle k^3\rangle -\langle k^2 \rangle^2} \,.
\end{equation}
Expression (\ref{r2}) is determined by the three first raw moments of $P(k)$, 
and also by  $B$ and $C$ that are the quantities embodying the information on the linear degree-degree 
correlations.   
 
For the Poisson distributed networks, by taking into account the analytical expressions for the moments of $P(k)$, 
it is straightforward to see that
\begin{equation} \label{rpoisson}
r= B -  \langle k \rangle + C \,.
\end{equation}
Clearly, dissassortative correlations are favored by vanishing $C$. 
Only for positive $r$ the growth of $C$ is convenient, but   
$B-\langle k \rangle$ can vary in a wider interval than $C$ (twice wider in this case).
The existence of two regimes in the increase of $C(r)$, observed in Fig.~\ref{fig:ERC}, is consistent with this picture.
In other words, Eq.~(\ref{rpoisson}) indicates that, in the rewiring process of a Poisson network 
to attain $r_\star$, 
as soon as $P(k)$ is conserved and $C$ remains very small, 
then $r$ is ruled predominantly  by $B$.

For the power-law distributed networks, some qualitatively similar effects occur, as far as the 
relation of $r$ with $B$ and $C$ is always linear and $C$ is constrained to a narrower interval than 
$B$. 
The formation of triangles also in this case contributes only for assortative values of $r$ (above the crossover), 
with values of $C$ larger than in the Poisson case but still small. 
Then also in this case the increase of the branching must drive rewiring. 
In contrast to the Poisson case, the other terms in Eq. (\ref{r2}), 
related to the moments of $P(k)$, might have a crucial influence on $r$  because 
of the divergence, in the infinite network limit, of the $n$th moments for $\gamma\ge n+1$.

Let us analyze the large $k_{max}$ (hence $N$)  limit, 
setting aside the marginal (logarithmic) cases. 
Here, we use the result $k_{max}\sim N$. However, if used instead 
$k_c \sim N^{1/(\gamma-1)}$, the conclusions would remain the same. 
Considering the expressions for the moments (e.g., Eq. (\ref{k1})),  
one has, for $3<\gamma<4$: $r \sim [B-{\cal O}(1)]/{\cal O}(N^{ 4-\gamma })$, 
while for $2<\gamma<3$: $r \sim [B-{\cal O}(N^{ 3-\gamma })]/{\cal O}(N )$, 
meaning ${\cal O}(x^\alpha) \sim a x^\alpha$, with $a>0$. 
To approach the lower limit  $r=-1$, one must have minimal $B$. 
If it is of order greater than the other terms in the numerator, then one can not have negative $r$, 
because $B$ is non-negative and it will dominate the numerator. 
Thus, negative $r$ can arise only if $B$ is of the same or lower order. 
But in that case $r\to 0$ in the large $N$ limit. 
This explains why the lower bound $r_{min}$ tends to 0 when $\gamma\le4$ 
(see Fig.~\ref{fig:rlim}(b)). 
Along this line, however, $r_{min}$ is not expected to vanish when $\gamma>4$, 
but to tend to a small finite value. 
Similarly, to attain a non-null upper bound of $r$, $B$ needs to grow like the denominator, 
otherwise, the upper bound will be negative and also vanish when $N\to \infty$, 
leading to the collapse of the upper bound too. However, this does not necessarily happens 
if $B$ is driven to grow enough during rewiring, which is what seems to be happen according to 
Fig.~\ref{fig:rlim}(b).    
 
The connection between $r$ and distance measures is not so direct analytically.  
Numerical results showed that, for networks with localized distribution of links,  
changing $r$  modifies significantly the mean path length only    
when correlations are assortative  ($r>0.5$) and   $\langle k\rangle$ small. 
These changes could be related to the induced fragmentation, that diminishes by increasing $\langle k\rangle$.  
Then, the impact of $r$ becomes less important as $\langle k\rangle$ increases.
Meanwhile, the influence on the diameter is more dramatic. 
In power-law networks,    the modification of the mean path length by $r$ 
is a bit more marked even if 
fragmentation is absent for $\langle k \rangle \ge 4$, while the diameter is not largely affected.
In both cases, the modification of characteristic lengths  
that occur when varying $r$ may affect transport processes  
and should be taken into account either when interpreting or designing 
numerical experiments on top of these networks.    

\section*{Acknowledgements:}

We acknowledge partial financial support from Brazilian  Agency CNPq.
The authors are grateful to professor Thadeu Penna for having provided 
the computational resources of the Group of Complex Systems of the 
Universidade Federal Fluminense, Brazil, where some of 
the simulations were performed.

\bibliographystyle{unsrt}

\end{document}